\documentclass[11pt]{article}

\usepackage[T1]{fontenc}
\usepackage[utf8]{inputenc}
\usepackage{lmodern}
\usepackage{microtype}
\usepackage[a4paper,margin=25mm]{geometry}
\usepackage{amsmath,amssymb,mathtools}
\usepackage{graphicx}
\usepackage{booktabs,tabularx,array,ragged2e}
\usepackage{enumitem}
\usepackage{caption}
\usepackage{float}
\usepackage{cite}
\usepackage[hidelinks]{hyperref}
\usepackage{xurl}

\DeclareUnicodeCharacter{2212}{\ensuremath{-}}
\renewcommand{\arraystretch}{1.08}

\title{CARE: A Responsibility-Oriented Architecture for Domain-Scoped Resolution in Modular and Upgradeable Smart Contracts}
\author{}
\date{}

\begin{document}

\begin{center}
{\LARGE\bfseries CARE: A Responsibility-Oriented Architecture for Domain-Scoped Resolution in Modular and Upgradeable Smart Contracts\par}

\vspace{1em}

{\large
Ali Rajabi Nekoo\textsuperscript{1},
Laleh Rasoul\textsuperscript{1},
Azadeh Zamanifar\textsuperscript{1,*}
\par}

\vspace{0.5em}

\textsuperscript{1}
Department of Computer Engineering,
SR.C, Islamic Azad University, Tehran, Iran

\vspace{0.3em}

\textit{Emails:}
ali.rajabinekoo@iau.ir,
laleh.rasoul@iau.ir,
azamanifar@iau.ac.ir

\end{center}

\begin{abstract}
Modular and upgradeable smart-contract systems enable replaceable execution logic, but modular dispatch alone does not define which components own architectural responsibilities such as coordination, resolution, shared state, reusable services, and observation. This paper presents CARE, a responsibility-oriented architecture that makes these ownership boundaries explicit and introduces domain-scoped capability resolution, separating execution coordination from the authority that selects an execution target. CARE is defined through a formal architectural model from which five structural properties are derived. Its behavioral semantics are further encoded in TLA+ and evaluated with TLC against seven architectural invariants. Intermediate TLC results were used to refine the architectural model before the final validation run. A controlled empirical evaluation against EIP-2535 and a routing-only baseline shows that CARE introduces an approximately fixed 7.7k-gas runtime routing premium in the evaluated workloads, while localized capability replacement remains close to the routing-only control and substantially less expensive than the evaluated EIP-2535 reconfiguration path. Adversarial tests further characterize authorization controls and residual risks associated with delegated execution and reentrancy. These results position CARE as an architectural model for explicit responsibility ownership and localized evolution rather than as a universal replacement for existing smart-contract upgradeability mechanisms.
\end{abstract}


\nocite{ref1, ref2, ref3, ref4, ref5, ref6, ref7, ref8, ref9, ref10, ref11, ref12, ref13, ref14, ref15, ref16, ref17, ref18, ref19, ref20, ref21, ref22, ref23}
\section{Introduction}\label{sec:introduction}

Smart contracts increasingly implement systems whose behavior extends beyond a small collection of independent functions. Decentralized marketplaces, asset-management systems, financial applications, games, identity platforms, and governance systems may contain many business capabilities, persistent state, administrative rules, reusable services, and components that evolve at different rates. As these systems grow, their architecture becomes a question not only of implementing correct behavior, but also of determining where responsibilities reside and how those responsibilities evolve.

Upgradeability mechanisms address an important part of this problem. Proxy-based architectures separate a stable interaction point from replaceable implementation logic, while modular mechanisms such as EIP-2535 allow multiple implementation components to coexist behind a common entry point. These approaches make code replacement and modular dispatch practical. They do not, however, inherently define how broader architectural responsibilities are distributed.

This distinction becomes important in systems containing several functional domains. A contract can be modular in terms of deployment units while still centralizing unrelated architectural decisions in one routing structure. The same component may receive requests, own the global routing table, select implementations, expose shared configuration, and participate in system-wide concerns. Individual implementations may be replaceable, yet ownership of the architectural decisions surrounding those implementations remains implicit.

We argue that execution modularity and responsibility modularity are distinct concerns.

A modular dispatch mechanism answers a question such as:

Which implementation should receive this invocation?

A responsibility-oriented architecture must additionally answer:

Who owns the decision, within which scope, and which other concerns remain independent?

CARE addresses this second problem.

CARE is a responsibility-oriented architecture for modular and upgradeable smart-contract systems. It separates six architectural responsibilities into explicit component classes:

\begin{itemize}
\item
  a Runtime Kernel (RK) coordinates execution;
\item
  a Resolution Authority (RA) owns execution resolution within one domain;
\item
  an Execution Module (EM) owns business behavior;
\item
  Shared Architectural State (SS) owns architectural state;
\item
  a Shared Architectural Service (SA) owns reusable shared behavior;
\item
  an Observation Hub (OH) owns architectural observation.
\end{itemize}

The central mechanism is domain-scoped capability resolution. Instead of maintaining one global capability-to-implementation decision space, an execution request identifies both a capability and a resolution domain. The Runtime Kernel coordinates the request but does not own the capability binding. The Resolution Authority for the selected domain determines the active Execution Module.

The normal execution path is therefore:

\[
\mathrm{Request} \rightarrow \mathrm{RK} \rightarrow \mathrm{RA}_d \rightarrow \mathrm{EM} \rightarrow \mathrm{Result}
\]

where \(\mathrm{RA}_d\) owns resolution for domain \(d\).

This design separates execution coordination from execution resolution. It also permits the same capability identifier to have different implementations in different domains without requiring one global capability namespace.

CARE is intentionally narrower than a complete smart-contract lifecycle framework. It does not define a universal governance protocol, version-management system, rollback mechanism, storage-migration framework, vulnerability detector, or implementation-specific security library. Those mechanisms may be composed with CARE. The architectural contribution is to define who owns the runtime responsibilities within which those mechanisms operate.

To make these boundaries precise, CARE is defined independently of Solidity through a formal architectural model. The model specifies component classes, responsibility ownership, legal dependency directions, domain membership, resolution functions, state-access rules, service relationships, and observation boundaries. Five structural properties follow from the model:

\begin{enumerate}
\def\labelenumi{\arabic{enumi}.}
\item
  Unique Responsibility Ownership;
\item
  Domain-Scoped Centralized Capability Resolution;
\item
  Architectural State Ownership Independence;
\item
  Shared Service Ownership Independence;
\item
  Structural Observation Separation.
\end{enumerate}

These properties characterize the architecture. They are not claims that every CARE implementation is automatically more maintainable, secure, reliable, or scalable.

A second validation layer encodes the behavioral semantics of CARE in TLA+. Seven architectural invariants are checked with TLC over a bounded finite-state model. Intermediate TLC results were used during model development to refine the architectural constraints. The final model completed exploration of 360,980 distinct reachable states without violation of the seven enabled invariants.

A Solidity reference implementation provides the third layer of evidence. Its operational behavior is evaluated using Direct execution, an EIP-2535-oriented Diamond implementation, and a minimal routing-only control. The experiments examine runtime dispatch, routing cardinality, domain partitioning, localized capability replacement, and independently enabled architectural concerns.

The empirical results reveal a trade-off rather than a universal performance advantage. The evaluated CARE runtime path introduces an approximately fixed 7.7k-gas premium relative to the EIP-2535 execution path. Finer domain decomposition increases one-time deployment and configuration cost. Conversely, localized CARE capability replacement remains within approximately 0.61--1.72\% of the minimal routing control across the tested replacement sizes and requires less gas than the evaluated EIP-2535 reconfiguration operation. CARE is therefore primarily intended for modular systems in which capabilities or functional domains are expected to evolve independently. For simple or rarely evolving contracts, a direct implementation or simpler proxy/routing architecture may be more appropriate. These comparisons characterize the tested implementations and should not be generalized into a claim that CARE is universally cheaper than EIP-2535.

Finally, a threat-model-driven evaluation examines implementation-level security boundaries. Authorization controls reject unauthorized architectural mutations in the reference implementation. At the same time, delegated Execution Modules execute through delegatecall and therefore belong to the Runtime Kernel's trusted computing base. CARE also does not inherently prevent reentrant or nested cross-domain execution.

The contributions of this work are:

\begin{enumerate}
\def\labelenumi{\arabic{enumi}.}
\item
  \textbf{A responsibility-oriented architecture for modular smart-contract systems.\\
  } CARE defines explicit owners for execution coordination, execution resolution, business execution, architectural state, shared services, and observation.
\item
  \textbf{A formal model of domain-scoped capability resolution.\\
  } The architecture is specified through entities, responsibilities, relations, mappings, and constraints from which five structural properties are derived.
\item
  \textbf{Machine-checked validation of selected architectural behavior.\\
  } A TLA+ model is checked against seven architectural invariants, with intermediate TLC results informing refinement of the architectural model before the final validation run.
\item
  \textbf{A controlled empirical evaluation.\\
  } A Solidity reference implementation is compared with EIP-2535 and a routing-only control across runtime, architectural-growth, localized-evolution, and concern-ablation experiments.
\item
  \textbf{An explicit characterization of implementation trust boundaries.\\
  } Adversarial tests distinguish enforced authorization controls from residual risks including trusted delegated code and reentrant execution.
\end{enumerate}

The remainder of the paper first positions CARE relative to existing smart-contract evolution and modularity work, then defines the architecture and formal model. Machine-checked validation is followed by the empirical methodology and results. The final sections examine security boundaries, architectural trade-offs, threats to validity, and future work.

\section{Background and Related Work}\label{sec:background}

\subsection{Smart-Contract Evolution and Modular Execution}\label{smart-contract-evolution-and-modular-execution}

Although deployed contract bytecode is immutable, deployed smart-contract systems continue to evolve. Studies of smart-contract development report recurring difficulties involving language constraints, testing, security, tooling, and upgradeability \cite{ref1, ref2}, while broader reviews characterize smart-contract engineering as a combination of evolving implementation practices and unresolved development challenges \cite{ref7}. Maintenance studies similarly show that contract evolution extends beyond corrective fixes and includes adaptation and internal improvement \cite{ref3}, \cite{ref20, ref21, ref22}.

Upgradeable architectures address this need by introducing indirection between a stable interaction surface and implementation logic. Empirical work has shown that upgradeable contract structures are used in practice \cite{ref4}, while developer studies identify additional complexity around initialization, storage compatibility, and interaction among upgradeable components \cite{ref23}.

EIP-2535 extends modular execution by routing function selectors to multiple facets behind a Diamond. This makes it an important comparison point for CARE: both architectures can support multiple replaceable execution components behind a stable entry point.

The architectural unit of resolution differs, however. Diamond-style dispatch is organized around selector-to-facet relationships. CARE organizes execution around a pair:

\begin{center}
\(\left(\mathrm{Capability},\mathrm{Domain}\right)\)
\end{center}

and assigns ownership of resolution for each domain to an explicit Resolution Authority.

This distinction is not an argument that selector-based routing is deficient. EIP-2535 and CARE address partially overlapping but different questions. EIP-2535 defines a modular dispatch and reconfiguration mechanism. CARE defines responsibility ownership and the architectural scope within which an execution target is resolved.

\subsection{State, Lifecycle, and Responsibility Boundaries}\label{state-lifecycle-and-responsibility-boundaries}

Persistent state often has a longer lifecycle than the executable logic that manipulates it. Work on Ethereum state emphasizes the complexity and importance of persistent storage \cite{ref8}, while empirical upgradeability studies similarly treat preservation of existing state as a central concern \cite{ref4, ref23}.

CARE does not introduce a new storage-migration protocol. Instead, it applies the distinction between behavior and state at the architectural level. Shared Architectural State owns architectural information independently from the lifecycle of individual Execution Modules. An EM may access state without becoming the owner of that state.

This distinction is summarized as:

\[
\operatorname{Access}(x) \not\Rightarrow \operatorname{Ownership}(x)
\]

Version management and governance form another separate concern. LEAGAN addresses decentralized revision and version management \cite{ref6}, while TRUST considers a broader governed lifecycle involving version traceability, rollback-related concerns, and provenance \cite{ref5}. CARE does not reproduce those mechanisms. A capability-binding update in CARE may be governed through an external governance mechanism, but the governance protocol itself is outside the architecture defined here.

Prior smart-contract design-pattern research also examines modularity, maintainability, control, and security \cite{ref9}. CARE adopts a stricter responsibility-oriented interpretation of modularity: components are distinguished not merely because they are separately deployed but because they own different architectural decisions.

In particular:

\[
\mathrm{ExecutionCoordination} \neq \mathrm{ExecutionResolution} \neq \mathrm{BehaviorExecution}
\]

The three responsibilities may cooperate during one request while remaining independently owned.

\subsection{Security, Observation, and Formal Reasoning}\label{security-observation-and-formal-reasoning}

Smart-contract security research includes defect classification, vulnerability taxonomies, weakness analysis, detection, and repair \cite{ref10, ref11, ref12, ref13, ref14, ref15}. Such work makes clear that architectural structure alone cannot establish implementation security.

CARE therefore does not infer security from modularity. Instead, its architecture makes trust and responsibility boundaries explicit, and the reference implementation is evaluated separately through adversarial tests.

Observation presents a similar distinction. Blockchain transactions and events provide visibility, but visible activity is not automatically architecturally understandable. Prior work distinguishes transparency, accountability, and understandability \cite{ref16}. CARE assigns architectural observation to a dedicated Observation Hub so that ownership of the observation concern is explicit.

This does not imply universal failure isolation. A telemetry failure may reasonably be ignored in one implementation, while a mandatory compliance record may need to make a transaction fail in another.

Formal methods have also been applied to smart contracts at multiple abstraction levels \cite{ref17}. Model-based approaches demonstrate the value of specifying relevant structures and properties before implementation \cite{ref18}. CARE combines two formal layers:

\[
\text{Architectural Formalization} \rightarrow \text{Behavioral Model Checking}
\]

The first defines the architecture deductively. The second uses TLA+ and TLC to explore selected transitions and invariants.

Neither layer constitutes bytecode-level verification.

Finally, the CARE model is conceptually independent of Solidity, but the empirical evidence in this paper is EVM-specific. Smart-contract platforms differ substantially in execution, state, and interaction semantics \cite{ref19}. A realization on another platform would therefore have to map CARE's responsibilities onto that platform rather than directly reproduce the Solidity implementation.

\begin{table}[H]
\centering
\caption{Positioning of CARE Relative to Existing Smart-Contract Evolution and Modularity Approaches}\label{tab:1}
\footnotesize
\renewcommand{\arraystretch}{1.15}
\setlength{\tabcolsep}{3pt}
\begin{tabularx}{\textwidth}{@{}>{\RaggedRight\arraybackslash}X>{\RaggedRight\arraybackslash}X>{\RaggedRight\arraybackslash}X>{\RaggedRight\arraybackslash}X@{}}
\toprule
\textbf{Approach} & \textbf{Primary concern} & \textbf{Main unit of change} & \textbf{CARE distinction} \\
\midrule
Proxy upgradeability \cite{ref4, ref23} & Replaceable implementation & Implementation & Adds domain ownership of resolution \\
EIP-2535 & Modular dispatch & Selector / facet & Separates coordination from domain-scoped resolution \\
State-oriented work \cite{ref8} & Persistent state & Storage / data & Defines independent architectural-state ownership \\
LEAGAN / TRUST \cite{ref5, ref6} & Version and lifecycle management & Version / lifecycle & Focuses on runtime responsibility organization \\
Design-pattern research \cite{ref9} & Reusable design knowledge & Pattern & Defines one responsibility-ownership model \\
Security research \cite{ref10,ref11,ref12,ref13,ref14,ref15} & Vulnerability analysis & Code / weakness & Characterizes trust boundaries rather than replacing security analysis \\
CARE & Responsibility-oriented runtime architecture & Capability / domain / responsibility & Explicit ownership plus domain-scoped resolution \\
\bottomrule
\end{tabularx}
\end{table}

The research gap addressed here is therefore not the absence of upgradeability, modularity, state management, governance, security, or formal methods individually. It is the absence, within the studied approaches, of a runtime architectural model combining explicit responsibility ownership with domain-scoped capability resolution while separately representing coordination, resolution, behavior, architectural state, reusable services, and observation.

\textbf{Open-source artifacts.}
The CARE artifacts are publicly available at
\url{https://github.com/rajabinekoo/CARE-Architecture},
including the reference implementation of the architecture, its formal
TLA+ specification, and the TLC model-checking results reported in this paper.

\section{CARE Architecture and Formal Model}\label{sec:architecture}

\subsection{Design Principles}\label{design-principles}

CARE is based on six architectural requirements.

\textbf{R1 --- Explicit responsibility ownership.}\\
Major architectural decisions should have identifiable owners rather than emerge implicitly from contract placement.

\textbf{R2 --- Domain-scoped resolution.}\\
Execution resolution should occur inside an explicit functional domain rather than depend on one global capability namespace.

\textbf{R3 --- Independent architectural-state ownership.}\\
Business modules may access architectural information without owning its lifecycle or mutation policy.

\textbf{R4 --- Independent shared-service ownership.}\\
Reusable capabilities should remain independent of individual workflows that consume them.

\textbf{R5 --- Explicit architectural observation.}\\
Observation should have an identifiable architectural owner rather than being implicitly distributed across business modules.

\textbf{R6 --- Localized evolution.}\\
Replacing one capability implementation should primarily modify resolution state for its domain rather than require unrelated modules or the central coordinator to change.

\begin{figure}[H]
\centering
\includegraphics[width=0.96\linewidth,height=0.70\textheight,keepaspectratio]{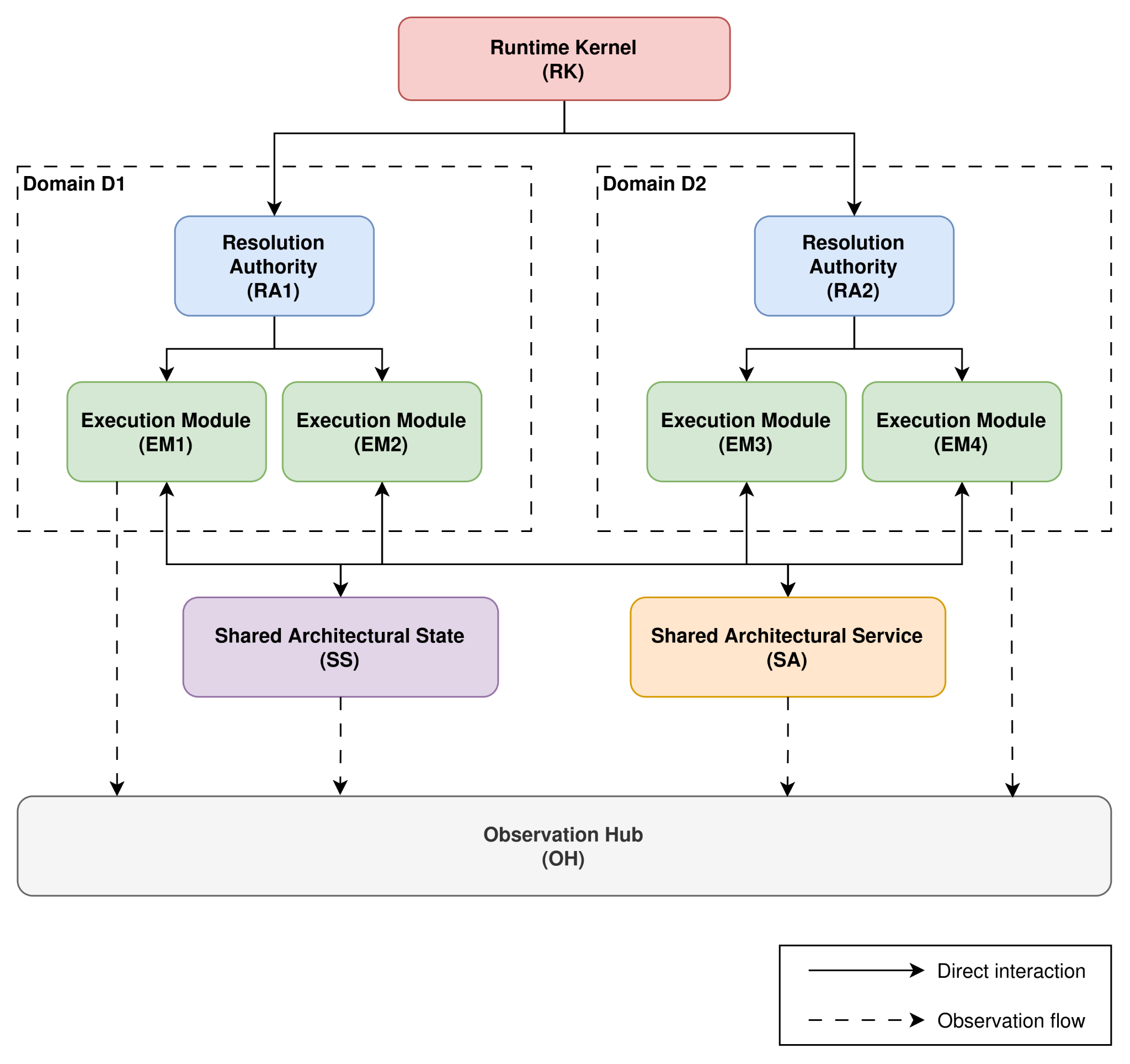}
\caption{High-level CARE architecture and responsibility boundaries across the Runtime Kernel, Resolution Authorities, Execution Modules, Shared Architectural State, Shared Architectural Services, and Observation Hub.}\label{fig:care-architecture}
\end{figure}

\subsection{Architectural Components}\label{architectural-components}

CARE contains six component classes.

\subsubsection{Runtime Kernel}\label{runtime-kernel}

The Runtime Kernel (RK) owns Execution Coordination.

It receives an execution request, identifies the requested domain and capability, locates the responsible Resolution Authority, obtains the current target, and transfers control to the selected Execution Module.

Its role is:

\[
\mathrm{Request} \rightarrow \mathrm{Coordination} \rightarrow \mathrm{Resolution} \rightarrow \mathrm{Execution}
\]

RK does not own business behavior and does not own the domain's capability-to-module resolution policy.

\subsubsection{Resolution Authority}\label{resolution-authority}

A Resolution Authority (RA) owns Execution Resolution for one domain.

Given a capability identifier \(c\), an RA determines which Execution Module currently implements that capability inside its domain.

Multiple RAs may exist simultaneously. Consequently, identical capability identifiers may resolve differently:

\[
\operatorname{Resolve}(C_1, \mathrm{RA}_1)=\mathrm{EM}_1
\]

\[
\operatorname{Resolve}(C_1, \mathrm{RA}_2)=\mathrm{EM}_3
\]

without ambiguity.

\subsubsection{Execution Module}\label{execution-module}

An Execution Module (EM) owns Behavior Execution.

An EM implements a business capability after resolution has selected it. It may consume architectural state or shared services, but use of those concerns does not transfer ownership to the module.

\subsubsection{Shared Architectural State}\label{shared-architectural-state}

Shared Architectural State (SS) owns Architectural State Management.

SS represents a logical state-ownership boundary. This does not require all architectural information to be physically stored inside a single contract. A platform may distribute the representation while preserving an identifiable architectural owner.

Operational business state is not automatically SS merely because several modules use it. SS refers specifically to architectural information whose lifecycle belongs to the architecture rather than to one business module.

\subsubsection{Shared Architectural Service}\label{shared-architectural-service}

A Shared Architectural Service (SA) owns Shared Capability Provision.

An SA exposes reusable behavior to multiple EMs while preserving its own responsibility and lifecycle boundary.

The relationship is usage:

\begin{center}    
$\operatorname{Use}(\mathrm{EM},\mathrm{SA})$
\end{center}

rather than ownership.

In the formal model evaluated here, SA does not depend on SS. This preserves an explicit separation between reusable behavior and architectural-state ownership.

\subsubsection{Observation Hub}\label{observation-hub}

The Observation Hub (OH) owns Architectural Observation.

RK, RA, EM, or SA may publish architectural observations through OH. The architecture specifies ownership of the observation concern, not a universal policy for observation failure.

\begin{table}[H]
\centering
\caption{CARE Architectural Components and Responsibility Ownership}\label{tab:2}
\footnotesize
\renewcommand{\arraystretch}{1.15}
\setlength{\tabcolsep}{3pt}
\begin{tabularx}{\textwidth}{@{}>{\RaggedRight\arraybackslash}X>{\RaggedRight\arraybackslash}X>{\RaggedRight\arraybackslash}X>{\RaggedRight\arraybackslash}X@{}}
\toprule
\textbf{Component} & \textbf{Primary responsibility} & \textbf{Typical interaction} & \textbf{Does not own} \\
\midrule
RK & Execution coordination & RA, EM, SS, OH & Resolution policy; business behavior \\
RA & Domain-scoped resolution & RK, OH & Business behavior; architectural state \\
EM & Behavior execution & RK, SS, SA, OH & Resolution; SS ownership \\
SS & Architectural-state management & RK, EM & Business behavior \\
SA & Shared capability provision & EM, OH & Workflow ownership; resolution \\
OH & Architectural observation & Observation producers & Coordination; resolution; execution \\
\bottomrule
\end{tabularx}
\end{table}

\subsection{Domain-Scoped Execution}\label{domain-scoped-execution}

Every execution request identifies:

\begin{enumerate}
\def\labelenumi{\arabic{enumi}.}
\item
  a capability \(c\);
\item
  the domain whose Resolution Authority owns the resolution decision.
\end{enumerate}

The normal execution path is:

\[
\mathrm{Req} \rightarrow \mathrm{RK} \rightarrow \mathrm{RA}_d \rightarrow \mathrm{EM} \rightarrow \mathrm{Result}
\]

The Runtime Kernel therefore owns the execution path but not the resolution mapping itself.

This becomes important during evolution.

Suppose:

\[
\operatorname{Resolve}(C_1,\mathrm{RA}_1)=\mathrm{EM}_1
\]

and the implementation of \(C_1\) in domain \(\mathrm{RA}_1\) is replaced by \(\mathrm{EM}_2\). The relevant transition is:

\[
\operatorname{Resolve}(C_1,\mathrm{RA}_1): \mathrm{EM}_1 \rightarrow \mathrm{EM}_2
\]

No corresponding change is implied for:

\[
\operatorname{Resolve}(C_1,\mathrm{RA}_2)
\]

or any other unrelated binding.

This is the structural basis of localized capability evolution.

\begin{figure}[H]
\centering
\includegraphics[width=0.96\linewidth,height=0.70\textheight,keepaspectratio]{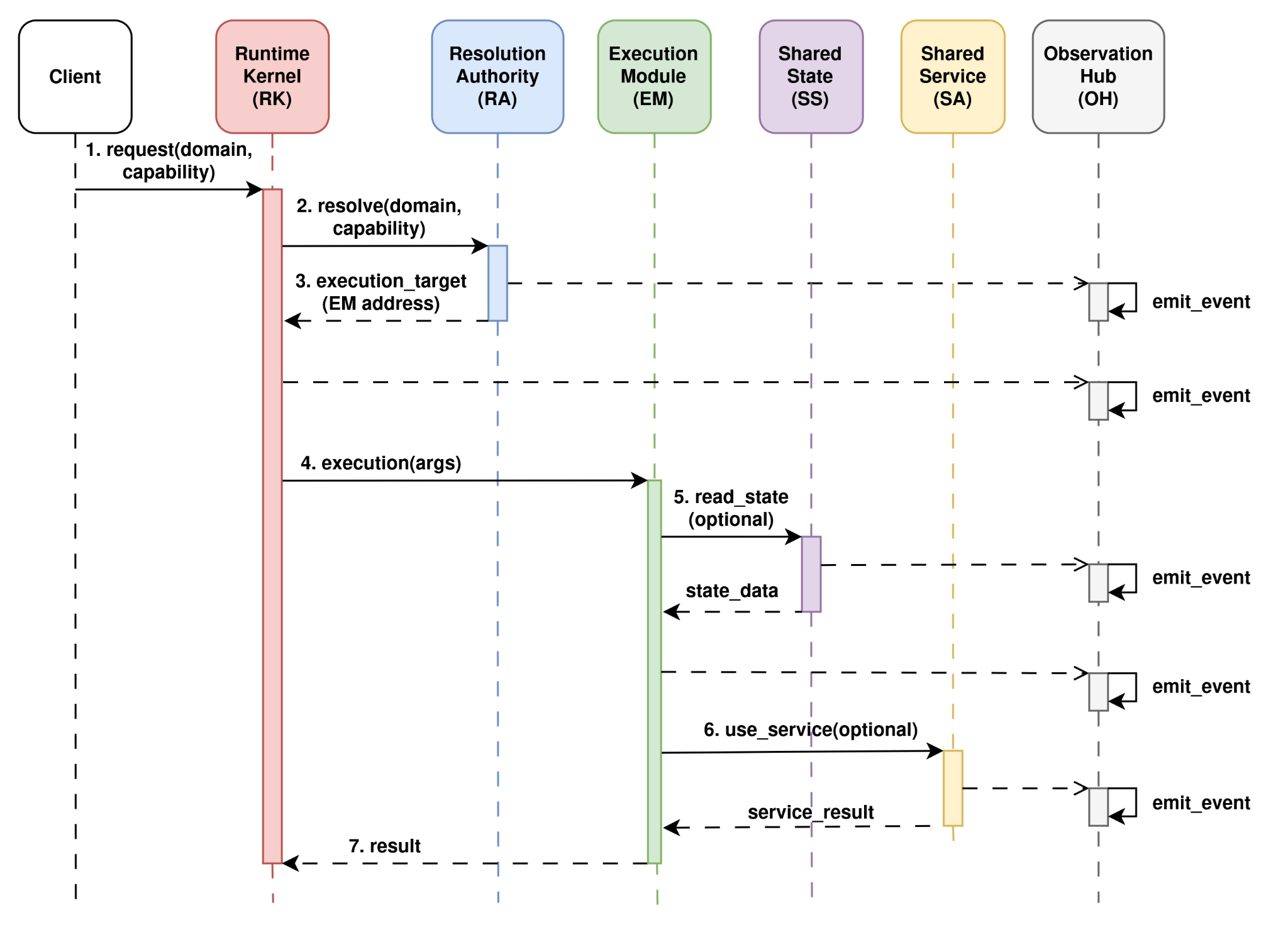}
\caption{Domain-scoped capability execution in CARE, showing the separation of execution coordination, domain-specific resolution, behavior execution, shared architectural concerns, and observation.}\label{fig:domain-scoped-execution}
\end{figure}

\subsection{Formal Definition}\label{formal-definition}

CARE is modeled as:

\[
A=(E,R,M,C)
\]

where:

\begin{itemize}
\item
  \(E\) is the set of architectural entities;
\item
  \(R\) is the set of architectural responsibilities;
\item
  \(M\) contains architectural relations and mappings;
\item
  \(C\) contains architectural constraints.
\end{itemize}

The entity set is partitioned into six pairwise-disjoint classes:

\[
E=\mathrm{RK}\cup \mathrm{RA}\cup \mathrm{EM}\cup \mathrm{SS}\cup \mathrm{SA}\cup \mathrm{OH}
\]

such that, for any distinct component classes \(X\) and \(Y\),

\[
X \cap Y = \varnothing
\]

The model contains one logical Runtime Kernel, architectural-state ownership boundary, and Observation Hub:

\[
|\mathrm{RK}|=1,\qquad |\mathrm{SS}|=1,\qquad |\mathrm{OH}|=1
\]

while allowing multiple Resolution Authorities and Execution Modules and optional Shared Architectural Services:

\[
|\mathrm{RA}|\geq 1,\qquad |\mathrm{EM}|\geq 1,\qquad |\mathrm{SA}|\geq 0
\]

The singleton constraints describe architectural roles in the model rather than requiring every realization to use exactly one physical contract for those roles.

\subsection{Responsibility Ownership}\label{responsibility-ownership}

The fundamental responsibility set is:

\[
R = \left\{ \begin{aligned} &\mathrm{ExecutionCoordination},\\ &\mathrm{ExecutionResolution},\\ &\mathrm{BehaviorExecution},\\ &\mathrm{StateManagement},\\ &\mathrm{SharedCapabilityProvision},\\ &\mathrm{ArchitecturalObservation} \end{aligned} \right\}
\]

Responsibility types are assigned as follows:

\[
\begin{aligned}
\operatorname{OwnerType}(\mathrm{ExecutionCoordination}) &= \mathrm{RK},\\
\operatorname{OwnerType}(\mathrm{ExecutionResolution}) &= \mathrm{RA},\\
\operatorname{OwnerType}(\mathrm{BehaviorExecution}) &= \mathrm{EM},\\
\operatorname{OwnerType}(\mathrm{StateManagement}) &= \mathrm{SS},\\
\operatorname{OwnerType}(\mathrm{SharedCapabilityProvision}) &= \mathrm{SA},\\
\operatorname{OwnerType}(\mathrm{ArchitecturalObservation}) &= \mathrm{OH}.
\end{aligned}
\]

Multiple instances may own independent instances of the same responsibility category. Multiple RAs, for example, each own resolution within their respective domains. The model prohibits ambiguity over ownership of one concrete architectural decision while permitting distributed ownership across independent domains.

\subsection{Invocation Constraints}\label{invocation-constraints}

CARE does not permit arbitrary architectural dependencies.

The legal relation is:

\[
\begin{aligned} \mathrm{InvokeRel} \subseteq {}& (\mathrm{RK}\times \mathrm{RA}) \cup(\mathrm{RK}\times \mathrm{EM}) \cup(\mathrm{RK}\times \mathrm{SS})\\ &\cup(\mathrm{EM}\times \mathrm{SS}) \cup(\mathrm{EM}\times \mathrm{SA})\\ &\cup(\mathrm{RK}\times \mathrm{OH}) \cup(\mathrm{RA}\times \mathrm{OH})\\ &\cup(\mathrm{EM}\times \mathrm{OH}) \cup(\mathrm{SA}\times \mathrm{OH}) \end{aligned}
\]

These edges represent permitted architectural dependency directions. They do not require each implementation to exercise every edge.

For example,

\[
\mathrm{EM}\rightarrow \mathrm{SA}
\]

is permitted, whereas

\[
\mathrm{SA}\rightarrow \mathrm{SS}
\]

is not part of the CARE model evaluated in this paper.

The architectural invocation graph satisfies:

\[
G(E,\mathrm{InvokeRel})\text{ is a DAG}
\]

This is an architectural dependency constraint. It must not be confused with runtime reentrancy: EVM calls may still produce nested execution despite an acyclic architectural dependency model.

\subsection{Domains and Resolution}\label{domains-and-resolution}

Every Execution Module belongs to one resolution domain:

\[
\operatorname{Domain}:\mathrm{EM}\rightarrow \mathrm{RA}
\]

Requests are mapped to a capability and Resolution Authority:

\[
\operatorname{ReqTarget}:\mathrm{Req}\rightarrow \mathrm{Cap}\times \mathrm{RA}
\]

If:

\begin{center}    
\(\operatorname{ReqTarget}(req)=(c,ra)\)
\end{center}

then \(ra\) owns resolution of capability \(c\) for that request.

Resolution is defined as:

\[
\operatorname{Resolve}:\mathrm{Cap}\times \mathrm{RA}\rightarrow \mathrm{EM}
\]

and is single-valued for an active capability-domain binding:

\[
\forall c\in \mathrm{Cap},\forall ra\in \mathrm{RA}: \left| \{em\mid \operatorname{Resolve}(c,ra)=em\} \right|\leq 1
\]

When a module is resolved, its domain must match the resolving authority:

\[
\operatorname{Resolve}(c,ra)=em \Rightarrow \operatorname{Domain}(em)=ra
\]

This authority--module conformance condition became particularly important during machine-checked validation.

\subsection{Architectural State, Services, and Observation}\label{architectural-state-services-and-observation}

State access is distinct from state ownership.

CARE defines:

\[
\mathrm{StateAccess} \subseteq (\mathrm{RK}\cup \mathrm{EM}\cup \mathrm{AdminUser})\times \mathrm{SS}
\]

and architectural reads as:

\[
\operatorname{StateRead}: (\mathrm{RK}\cup \mathrm{EM})\times \mathrm{SS}\times \mathrm{Ctx} \rightarrow \mathrm{State}
\]

An Execution Module may therefore consume SS information without owning its lifecycle.

Mutation is more restricted:

\[
\operatorname{StateWrite}: \Omega\times \mathrm{SS}\times \mathrm{Ctx} \rightarrow \mathrm{new\_State}
\]

where the architectural model permits:

\[
\Omega=\mathrm{RK}\cup \mathrm{AdminUser}
\]

The Solidity reference implementation adopts a stricter policy and restricts SS mutation to its administrative authority.

CARE therefore distinguishes:

\[
\text{Read Access} \neq \text{Write Authority} \neq \text{State Ownership}
\]

Shared-service use is represented by:

\[
\operatorname{Use}:\mathrm{EM}\times \mathrm{SA}\rightarrow \mathrm{Cap}
\]

with:

\[
\operatorname{Use}(em,sa)\not\Rightarrow \operatorname{Own}(em,sa)
\]

Observation is separately owned by OH. Execution components may produce observations, but producing an observation does not make them owners of the observation mechanism.

\subsection{Derived Structural Properties}\label{derived-structural-properties}

The preceding definitions produce five architectural properties.

\subsubsection{DP1 --- Unique Responsibility Ownership}\label{dp1-unique-responsibility-ownership}

Each fundamental architectural responsibility has an explicit owner type, while instance-level ownership is scoped to the relevant component or domain.

The property eliminates ambiguity in the model over which architectural role owns a decision.

It does not prove lower maintenance effort.

\subsubsection{DP2 --- Domain-Scoped Centralized Capability Resolution}\label{dp2-domain-scoped-centralized-capability-resolution}

Within one domain, capability resolution is owned by one Resolution Authority:

\[
(c,ra)\rightarrow em
\]

Resolution is therefore centralized inside each domain while distributed across domains.

It does not establish unrestricted scalability.

\subsubsection{DP3 --- Architectural State Ownership Independence}\label{dp3-architectural-state-ownership-independence}

Access by an EM does not transfer state ownership:

\[
\operatorname{StateRead}(em,ss,ctx) \not\Rightarrow \operatorname{OwnerType}(\mathrm{StateManagement})=\mathrm{EM}
\]

Behavior modules can therefore evolve independently from ownership of architectural state.

This does not solve arbitrary data migration.

\subsubsection{DP4 --- Shared Service Ownership Independence}\label{dp4-shared-service-ownership-independence}

For:

\[
\operatorname{Use}(em_1,sa) \land \operatorname{Use}(em_2,sa)
\]

neither consumer becomes the architectural owner of \(sa\).

This creates an independent lifecycle boundary for reusable behavior.

\subsubsection{DP5 --- Structural Observation Separation}\label{dp5-structural-observation-separation}

Architectural observation is owned independently from execution coordination, resolution, and business execution.

This property is structural. It does not imply:

\[
\mathrm{ObservationFailure} \Rightarrow \mathrm{NoEffectOnExecution}
\]

for every implementation.

\begin{table}[H]
\centering
\caption{Derived CARE Structural Properties and Explicit Non-Claims}\label{tab:3}
\small
\renewcommand{\arraystretch}{1.15}
\setlength{\tabcolsep}{3pt}
\begin{tabularx}{\textwidth}{@{}>{\RaggedRight\arraybackslash}X>{\RaggedRight\arraybackslash}X>{\RaggedRight\arraybackslash}X@{}}
\toprule
\textbf{Property} & \textbf{Architectural statement} & \textbf{Not claimed} \\
\midrule
DP1 & Responsibilities have explicit owners & Proven maintainability \\
DP2 & Resolution is unique within a domain & Universal scalability \\
DP3 & SS ownership is independent of EM access & Automatic storage migration \\
DP4 & SA ownership is independent of consumers & Automatic service quality/reuse \\
DP5 & Observation is structurally separated & Universal failure isolation \\
\bottomrule
\end{tabularx}
\end{table}

The formal model therefore defines architectural structure and ownership. It does not establish freedom from implementation bugs, reentrancy, storage collision, malicious code, or platform-specific failures.

These concerns motivate the two additional evidence layers used below: machine-checked behavioral validation and executable evaluation.

\section{Machine-Checked Behavioral Validation}\label{sec:validation}

\subsection{Validation Model}\label{validation-model}

The core behavioral semantics of CARE were encoded in TLA+ to examine whether selected architectural invariants remain preserved across sequences of configuration and execution operations.

The specification follows:

\[
\mathrm{Spec} = \mathrm{Init} \land \Box[\mathrm{Next}]_{\mathrm{vars}}
\]

where \(\mathrm{Init}\) defines the valid initial configuration and \(\mathrm{Next}\) defines allowed transitions.

The modeled transitions include:

\begin{itemize}
\item
  domain registration;
\item
  capability binding;
\item
  architectural-state mutation;
\item
  domain-scoped execution;
\item
  capability reconfiguration;
\item
  the handled observation-failure behavior used by the reference policy.
\end{itemize}

TLC is used to exhaustively explore the reachable state space of a bounded finite instance. Intermediate TLC results were used during development to refine the architectural model and corresponding implementation constraints before the final validation run.

\begin{figure}[H]
\centering
\includegraphics[width=0.96\linewidth,height=0.70\textheight,keepaspectratio]{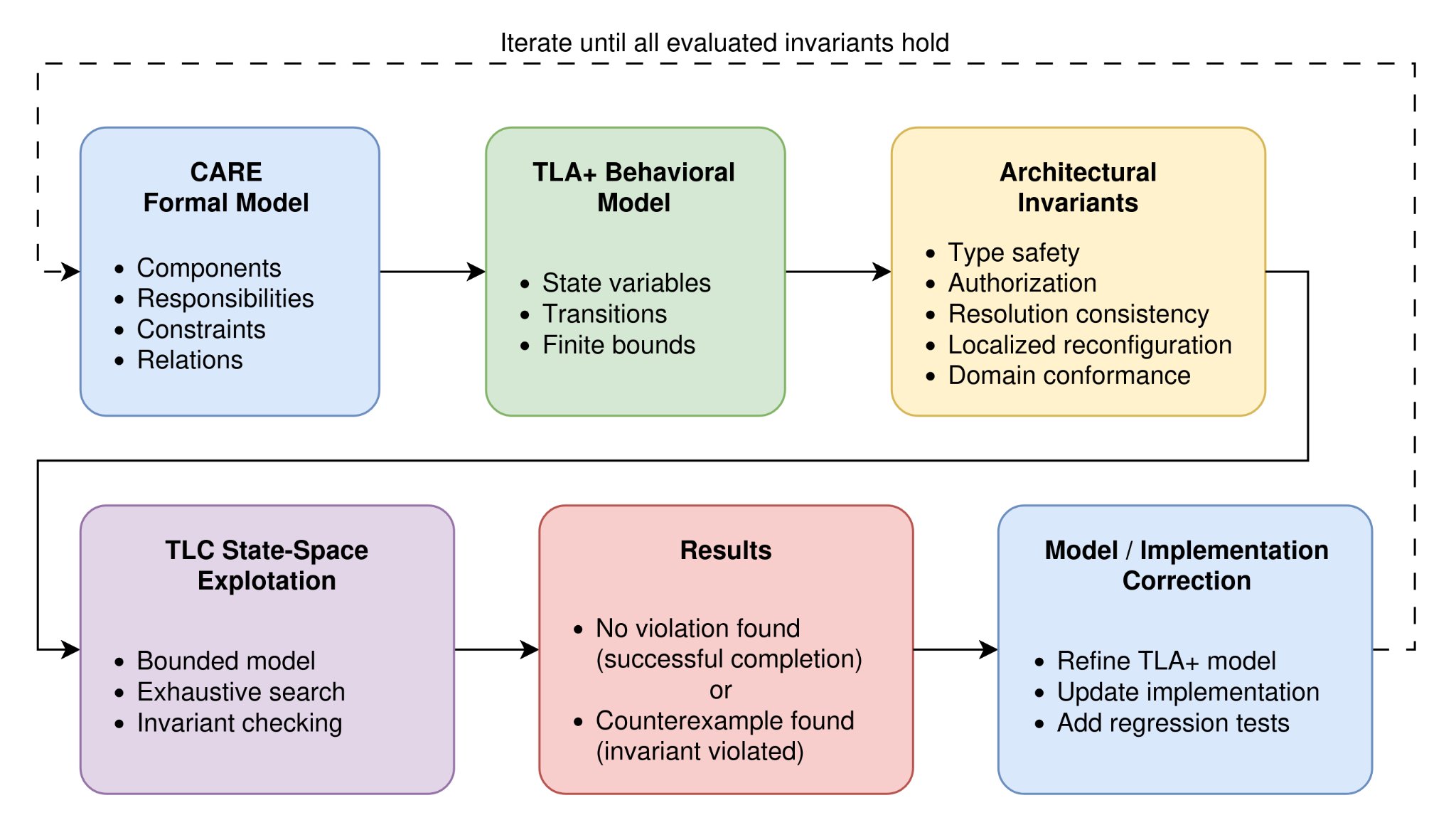}
\caption{Machine-checked validation workflow for CARE, from the formal architectural model to the bounded TLA+ behavioral specification, invariant checking with TLC, and model or implementation refinement.}\label{fig:validation-workflow}
\end{figure}

\subsection{Finite Bounds}\label{finite-bounds}

\begin{table}[H]
\centering
\caption{Finite Bounds of the TLA+ Behavioral Validation Model}\label{tab:4}
\small
\renewcommand{\arraystretch}{1.15}
\setlength{\tabcolsep}{3pt}
\begin{tabularx}{\textwidth}{@{}>{\RaggedRight\arraybackslash}X>{\RaggedRight\arraybackslash}X@{}}
\toprule
\textbf{Model element} & \textbf{Bound} \\
\midrule
Domains & D1, D2 \\
Resolution Authorities & RA1, RA2 \\
Capabilities & C1, C2 \\
Execution Modules & EM1, EM2, EM3 \\
Actors & admin, user, attacker \\
Architectural-state keys & K1, K2 \\
Architectural-state values & 0..1 \\
\bottomrule
\end{tabularx}
\end{table}

RA1 is associated with D1 and RA2 with D2 through an explicit authority-domain relation.

The model tracks domain registrations, capability bindings, architectural-state values, execution domain and capability, selected module, business status, observation status, selected previous configuration values, and the actor responsible for mutations.

The finite bounds exist to make exhaustive exploration tractable. They do not constrain the number of entities permitted by the CARE architecture.

\subsection{Architectural Invariants}\label{architectural-invariants}

Seven invariants were enabled in the final configuration.

\begin{table}[H]
\centering
\caption{Architectural Invariants Evaluated with TLC}\label{tab:5}
\small
\renewcommand{\arraystretch}{1.15}
\setlength{\tabcolsep}{3pt}
\begin{tabularx}{\textwidth}{@{}>{\RaggedRight\arraybackslash}X>{\RaggedRight\arraybackslash}X@{}}
\toprule
\textbf{Invariant} & \textbf{Evaluated property} \\
\midrule
TypeOK & Variables remain within their modeled domains \\
AuthorizedArchitecturalWriters & Configuration and SS mutation are performed only by the modeled administrator \\
LocalizedCapabilityReconfiguration & Updating one capability leaves unrelated bindings unchanged \\
ObservationFailureIsolation & Under the modeled handled-failure policy, an OH failure does not convert successful business execution into failure \\
ExecutionTargetWasResolved & A successful execution target follows the active domain $\rightarrow$ RA $\rightarrow$ capability $\rightarrow$ EM chain \\
UniqueDomainScopedResolution & Each domain-capability pair has at most one active execution target \\
RegisteredAuthorityMatchesDomain & A registered RA matches its declared domain \\
\bottomrule
\end{tabularx}
\end{table}

ObservationFailureIsolation is intentionally implementation-policy-specific. It is included because the evaluated Solidity reference implementation uses handled observation failures. It is not elevated to a universal CARE invariant.

\subsection{Exploration Results}\label{exploration-results}

The final model was evaluated using TLC 2.20 with breadth-first exploration and eight workers.

\begin{table}[H]
\centering
\caption{Final TLC State-Space Exploration Results}\label{tab:6}
\small
\renewcommand{\arraystretch}{1.15}
\setlength{\tabcolsep}{3pt}
\begin{tabularx}{\textwidth}{@{}>{\RaggedRight\arraybackslash}X>{\RaggedRight\arraybackslash}X@{}}
\toprule
\textbf{Metric} & \textbf{Result} \\
\midrule
Generated states & 1,805,161 \\
Distinct reachable states & 360,980 \\
States left on queue & 0 \\
Complete graph depth & 12 \\
Evaluated invariants & 7 \\
Recorded runtime & 9,872 ms \\
Outcome & No invariant violation found \\
\bottomrule
\end{tabularx}
\end{table}

Zero states remaining in the queue indicates that TLC completed exploration of the reachable state space for the selected finite bounds.

The appropriate conclusion is therefore:

\begin{quote}
No violation of the seven evaluated invariants was found in the explored bounded CARE model.
\end{quote}

This is machine-checked evidence for the modeled properties. It is not complete verification of Solidity source code, arbitrary EVM behavior, or unbounded CARE deployments.

The three evidence layers used by this paper can therefore be summarized as:

\[
\text{Formal Architecture} \rightarrow \text{Machine-Checked Behavior} \rightarrow \text{Executable Evaluation}
\]

\section{Evaluation Methodology}\label{sec:methodology}

\subsection{Research Questions}\label{research-questions}

The empirical evaluation examines the operational cost of the architectural decisions introduced by CARE.

Four research questions are considered.

\textbf{RQ1 --- Runtime Cost.}\\
What runtime cost does domain-scoped resolution introduce relative to the evaluated modular baselines?

\textbf{RQ2 --- Architectural Growth.}\\
How do routing cardinality and finer domain partitioning affect dispatch cost and one-time architectural setup cost?

\textbf{RQ3 --- Localized Evolution.}\\
What is the operational cost of replacing capability bindings, and how does it compare with a minimal routing update and the evaluated EIP-2535 reconfiguration path?

\textbf{RQ4 --- Independent Architectural Concerns.}\\
What incremental cost is introduced by SS, SA, and OH?

The purpose is not to select scenarios favorable to CARE. The experiments deliberately distinguish runtime execution, architectural construction, and evolution because the architecture may increase cost in one dimension while localizing it in another.

\subsection{Reference Implementation and Baselines}\label{reference-implementation-and-baselines}

The reference artifact implements the six CARE component classes in Solidity.

The Runtime Kernel receives a domain and capability, obtains the relevant Resolution Authority, resolves the corresponding Execution Module, and invokes it through delegated execution.

Three execution levels are used.

\subsubsection{Direct}\label{direct}

The Direct path executes the benchmark workload without modular routing. It provides a lower-bound reference for workload cost and is not treated as an upgradeable competitor.

\subsubsection{EIP-2535 Diamond}\label{eip-2535-diamond}

The practical Diamond baseline is a standards-oriented implementation. In benchmark v2.1, the evaluated Diamond stack contains:

\begin{itemize}
\item
  a DiamondCutFacet, installed during Diamond deployment and recorded through a DiamondCut event;
\item
  a DiamondLoupeFacet, installed before measured workload registration, dispatch, and reconfiguration;
\item
  Add, Replace, and Remove selector-to-facet bookkeeping through diamondCut;
\item
  standardized facet and selector introspection through the loupe interface;
\item
  fallback-based delegated execution.
\end{itemize}

The purpose of this baseline is to represent the management and dispatch mechanisms relevant to the evaluated EIP-2535 path rather than a selector-only proxy.

\subsubsection{DiamondCore}\label{diamondcore}

DiamondCore is a deliberately minimal selector-to-target routing map.

It is not an EIP-2535 implementation. Its purpose is to separate the cost of a simple mapping replacement from the additional facet-management work performed by the EIP-2535-oriented baseline.

This distinction is particularly important for RQ3.

\begin{figure}[H]
\centering
\includegraphics[width=0.96\linewidth,height=0.70\textheight,keepaspectratio]{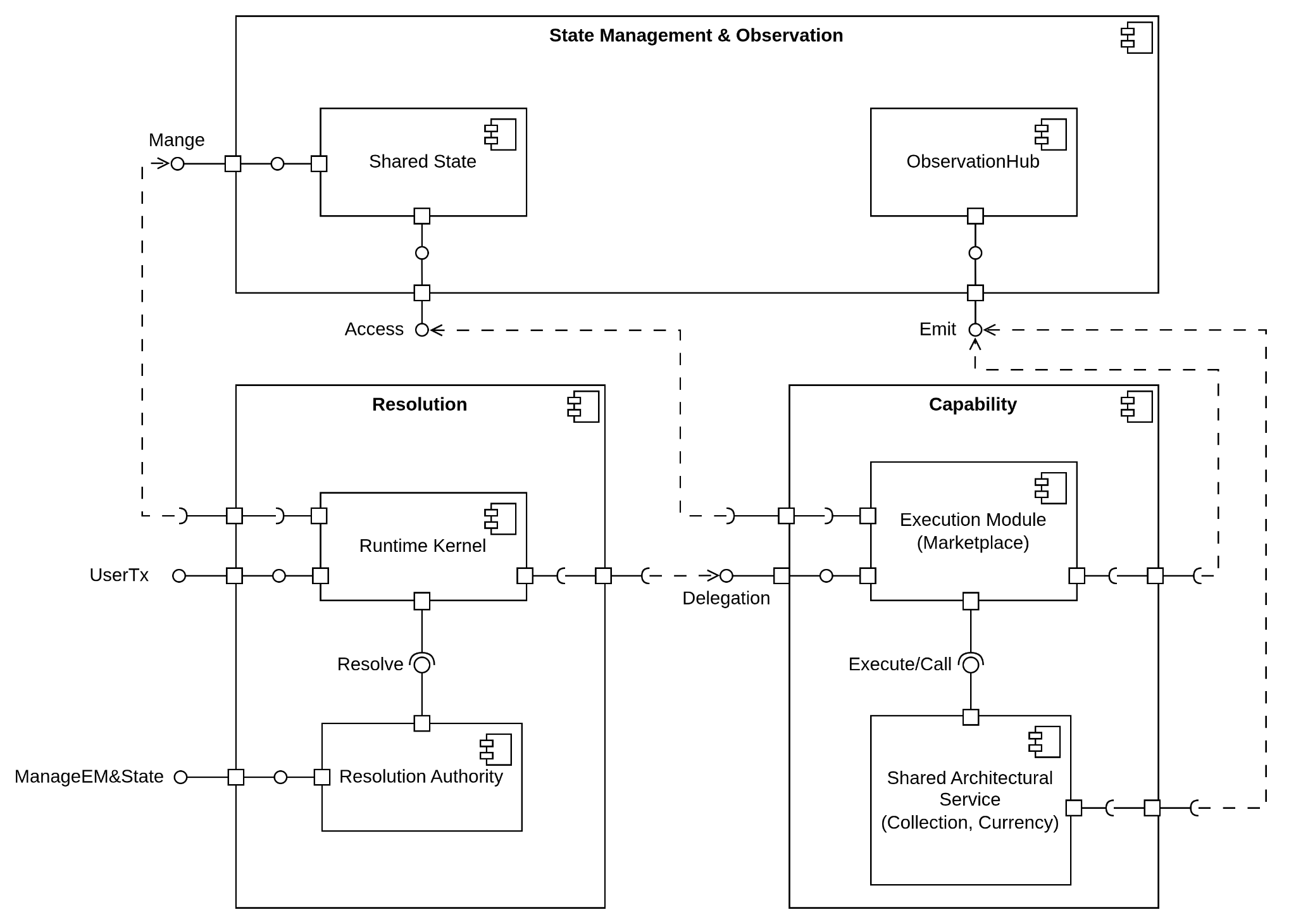}
\caption{Structure of the CARE Solidity reference implementation and the principal interaction paths among the Runtime Kernel, Resolution Authority, Execution Modules, Shared Architectural State, Shared Architectural Services, and Observation Hub.}\label{fig:reference-implementation}
\end{figure}

\begin{figure}[H]
\centering
\includegraphics[width=0.96\linewidth,height=0.70\textheight,keepaspectratio]{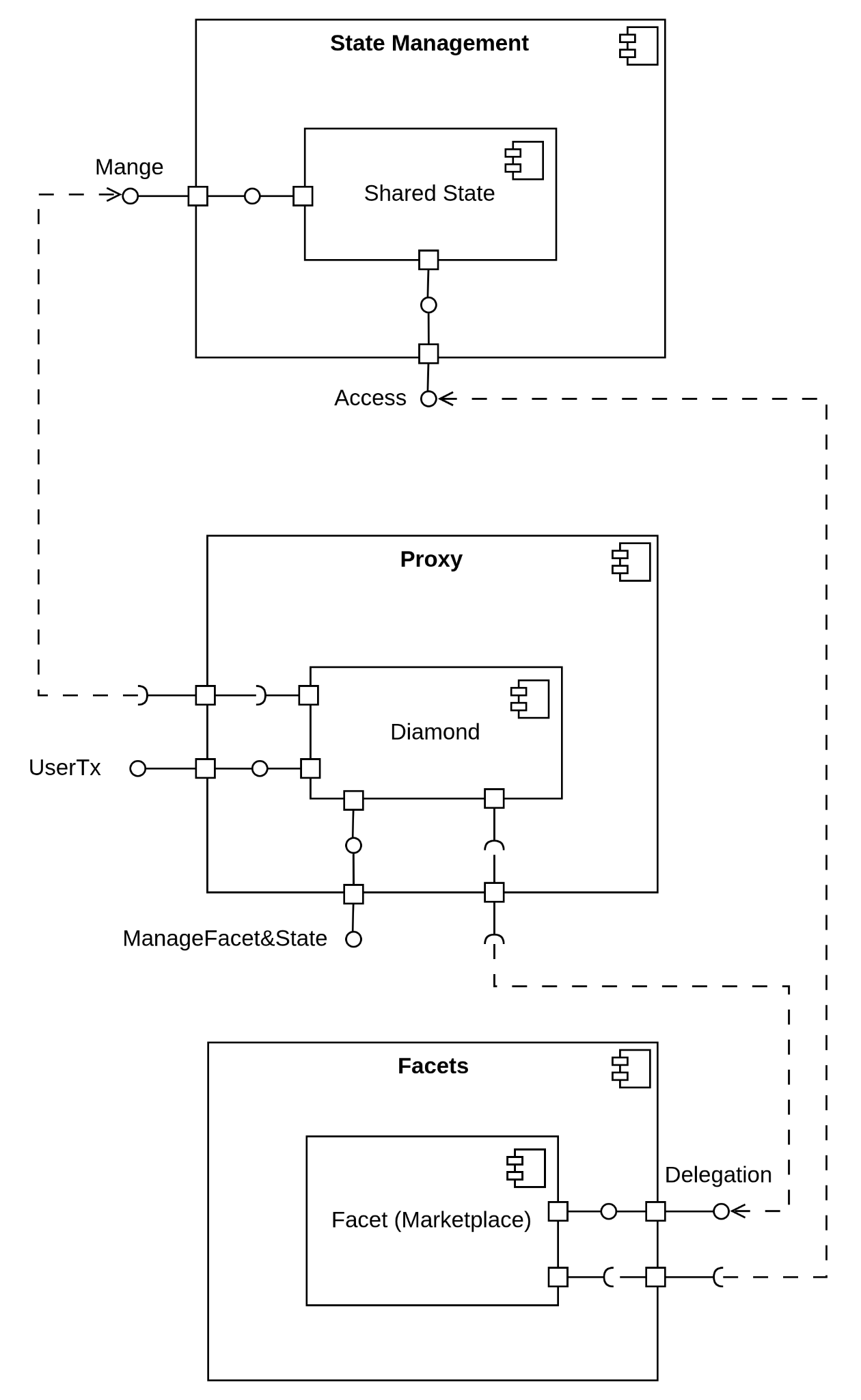}
\caption{EIP-2535 Diamond baseline used in the empirical evaluation, including DiamondCutFacet, DiamondLoupeFacet, selector-to-facet bookkeeping, and fallback-based delegated execution to the benchmark facet.}\label{fig:diamond-baseline}
\end{figure}

\subsection{Experimental Design}\label{experimental-design}

Five experiment groups are used.

\subsubsection{E1 --- Workload Ladder}\label{e1-workload-ladder}

The same benchmark behavior is exercised through Direct, EIP-2535, DiamondCore where applicable, and CARE paths.

\begin{center}
\small
\renewcommand{\arraystretch}{1.15}
\setlength{\tabcolsep}{3pt}
\begin{tabularx}{\textwidth}{@{}>{\RaggedRight\arraybackslash}X>{\RaggedRight\arraybackslash}X@{}}
\toprule
\textbf{Workload} & \textbf{Description} \\
\midrule
W0 & Minimal computation \\
W1 & Operational state write \\
W2 & External read \\
W3 & External state-changing call, state write, and event \\
W4 & Marketplace-style ERC-20/ERC-721 workflow \\
\bottomrule
\end{tabularx}
\end{center}

The workload ladder distinguishes absolute routing overhead from its relative contribution to transactions of different base costs.

\subsubsection{E2 --- Routing Cardinality}\label{e2-routing-cardinality}

Routing tables containing:

\[
1,\ 10,\ 50,\ 100
\]

entries are evaluated.

This experiment asks only whether dispatch gas changes within the tested range. It does not test unrestricted scalability.

\subsubsection{E3 --- Domain Growth}\label{e3-domain-growth}

A fixed total of 100 capabilities is partitioned across:

\[
1,\ 2,\ 5,\ 10,\ 20
\]

Resolution Authorities.

Capabilities per RA therefore become:

\[
100,\ 50,\ 20,\ 10,\ 5
\]

respectively.

Keeping total functionality fixed isolates the cost of architectural partitioning.

\subsubsection{E4 --- Evolution Sensitivity}\label{e4-evolution-sensitivity}

The benchmark replaces:

\[
1,\ 5,\ 10,\ 25
\]

routing entries using:

\begin{itemize}
\item
  EIP-2535;
\item
  DiamondCore;
\item
  CARE Resolution Authority bindings.
\end{itemize}

\subsubsection{E5 --- Architectural Concern Ablation}\label{e5-architectural-concern-ablation}

CARE-specific concerns are introduced incrementally.

\begin{center}
\small
\renewcommand{\arraystretch}{1.15}
\setlength{\tabcolsep}{3pt}
\begin{tabularx}{\textwidth}{@{}>{\RaggedRight\arraybackslash}X>{\RaggedRight\arraybackslash}X@{}}
\toprule
\textbf{Scenario} & \textbf{Enabled behavior} \\
\midrule
A0 & RK + RA + EM \\
A1 & A0 + SS read \\
A2 & A0 + SA invocation \\
A3 & A0 + SS + SA \\
A4 & A3 + OH \\
A5 & A3 + forced OH failure under reference handling policy \\
\bottomrule
\end{tabularx}
\end{center}

SS and SA are tested independently because they represent different architectural responsibilities.

\subsection{Measurements and Environment}\label{measurements-and-environment}

Gas consumption is the primary quantitative metric.

Gas is deterministic for equivalent transactions under the same EVM state, calldata, compiler configuration, and execution environment. Repeated identical executions are therefore not treated as independent statistical samples.

Selected opcode traces are additionally used to explain measured differences, particularly additional storage accesses, external calls, and delegated execution.

The evaluation artifact uses:

\begin{itemize}
\item
  benchmark version 2.0.0;
\item
  Solidity 0.8.28;
\item
  compiler optimization with 200 runs;
\item
  Cancun EVM target;
\item
  local Hardhat network;
\item
  Node.js v24.18.0; and
\item
  chain ID 31337.
\end{itemize}

The environment is intended for deterministic architectural comparison rather than network-level performance benchmarking.

\section{Results}\label{sec:results}

\subsection{RQ1 --- Runtime Cost}\label{rq1-runtime-cost}

\begin{table}[H]
\centering
\caption{Runtime Gas Consumption Across the Workload Ladder}\label{tab:7}
\scriptsize
\renewcommand{\arraystretch}{1.15}
\setlength{\tabcolsep}{3pt}
\begin{tabularx}{\textwidth}{@{}>{\RaggedRight\arraybackslash}X>{\RaggedRight\arraybackslash}X>{\RaggedRight\arraybackslash}X>{\RaggedRight\arraybackslash}X>{\RaggedRight\arraybackslash}X>{\RaggedRight\arraybackslash}X>{\RaggedRight\arraybackslash}X@{}}
\toprule
\textbf{Workload} & \textbf{Direct} & \textbf{EIP-2535} & \textbf{DiamondCore} & \textbf{CARE} & \textbf{CARE $-$ EIP-2535} & \textbf{Relative difference} \\
\midrule
W0 & 21,621 & 26,564 & 26,667 & 34,314 & 7,750 & 29.17\% \\
W1 & 43,656 & 48,599 & 48,702 & 56,361 & 7,762 & 15.97\% \\
W2 & 25,221 & 30,167 & 30,270 & 37,929 & 7,762 & 25.73\% \\
W3 & 70,619 & 75,565 & 75,668 & 83,327 & 7,762 & 10.27\% \\
W4 & 66,180 & 71,153 & 71,256 & 78,840 & 7,687 & 10.80\% \\
\bottomrule
\end{tabularx}
\end{table}

CARE consumed more gas than the EIP-2535 execution path in all five workloads. The absolute difference remained narrow:

\[
7{,}687 \leq \Delta \mathrm{Gas} \leq 7{,}762
\]

with an arithmetic mean of approximately 7,745 gas across the five workloads. The percentage difference changes because each workload has a different underlying execution cost. CARE's 29.17\% premium for W0 falls to 10.27\% for W3 and 10.80\% for W4. This does not mean that CARE becomes computationally faster as workload complexity increases; rather, the approximately fixed routing premium becomes a smaller fraction of a larger transaction.

Opcode traces support this interpretation. CARE performs one additional SLOAD and one additional STATICCALL relative to the evaluated EIP-2535 path, while both paths use one DELEGATECALL. This pattern is consistent with the additional domain-to-authority resolution step.

RQ1: the evaluated CARE implementation adds an approximately fixed 7.7k-gas runtime resolution premium.

\begin{figure}[H]
\centering
\includegraphics[width=0.96\linewidth,height=0.70\textheight,keepaspectratio]{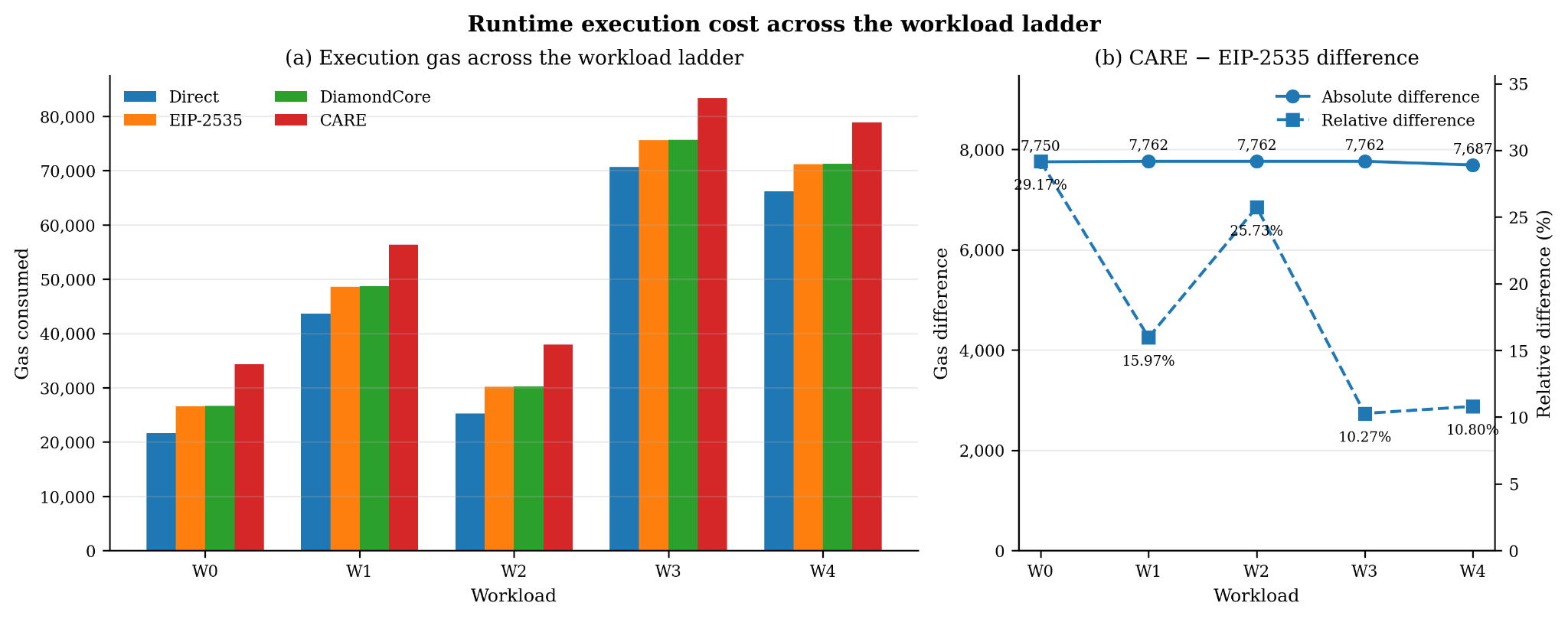}
\caption{Runtime gas consumption across the workload ladder for Direct execution, EIP-2535, DiamondCore, and CARE.}\label{fig:runtime-gas}
\end{figure}

\subsection{RQ2 --- Architectural Growth}\label{rq2-architectural-growth}

Routing cardinality had little observable effect within the tested range.

At 1, 10, 50, and 100 entries, the evaluated EIP-2535 path remained at approximately 26.6k gas, while CARE remained approximately 34.4k gas.

The domain-partitioning experiment produced a different effect.

\begin{table}[H]
\centering
\caption{CARE Domain Partitioning with 100 Total Capabilities}\label{tab:8}
\scriptsize
\renewcommand{\arraystretch}{1.15}
\setlength{\tabcolsep}{3pt}
\begin{tabularx}{\textwidth}{@{}>{\RaggedRight\arraybackslash}X>{\RaggedRight\arraybackslash}X>{\RaggedRight\arraybackslash}X>{\RaggedRight\arraybackslash}X>{\RaggedRight\arraybackslash}X>{\RaggedRight\arraybackslash}X@{}}
\toprule
\textbf{CARE domains} & \textbf{Capabilities / domain} & \textbf{EIP-2535 dispatch} & \textbf{CARE dispatch} & \textbf{Runtime difference} & \textbf{CARE partitioning cost} \\
\midrule
1 & 100 & 26,629 & 34,391 & 7,762 & 2,956,705 \\
2 & 50 & 26,629 & 34,391 & 7,762 & 3,356,418 \\
5 & 20 & 26,629 & 34,391 & 7,762 & 4,555,557 \\
10 & 10 & 26,629 & 34,391 & 7,762 & 6,554,098 \\
20 & 5 & 26,629 & 34,391 & 7,762 & 10,550,868 \\
\bottomrule
\end{tabularx}
\end{table}

Per-request dispatch remains unchanged as the number of configured RAs increases from one to twenty in this experiment.

One-time architecture construction does not.

CARE partitioning cost rises from approximately 2.96 million gas for one Resolution Authority to 10.55 million gas for twenty.

The increase results from deploying and registering additional authorities while the total number of capabilities remains fixed.

RQ2: finer domain decomposition did not increase measured request dispatch cost within the evaluated range, but it substantially increased one-time deployment and configuration cost.

\begin{figure}[H]
\centering
\includegraphics[width=0.96\linewidth,height=0.70\textheight,keepaspectratio]{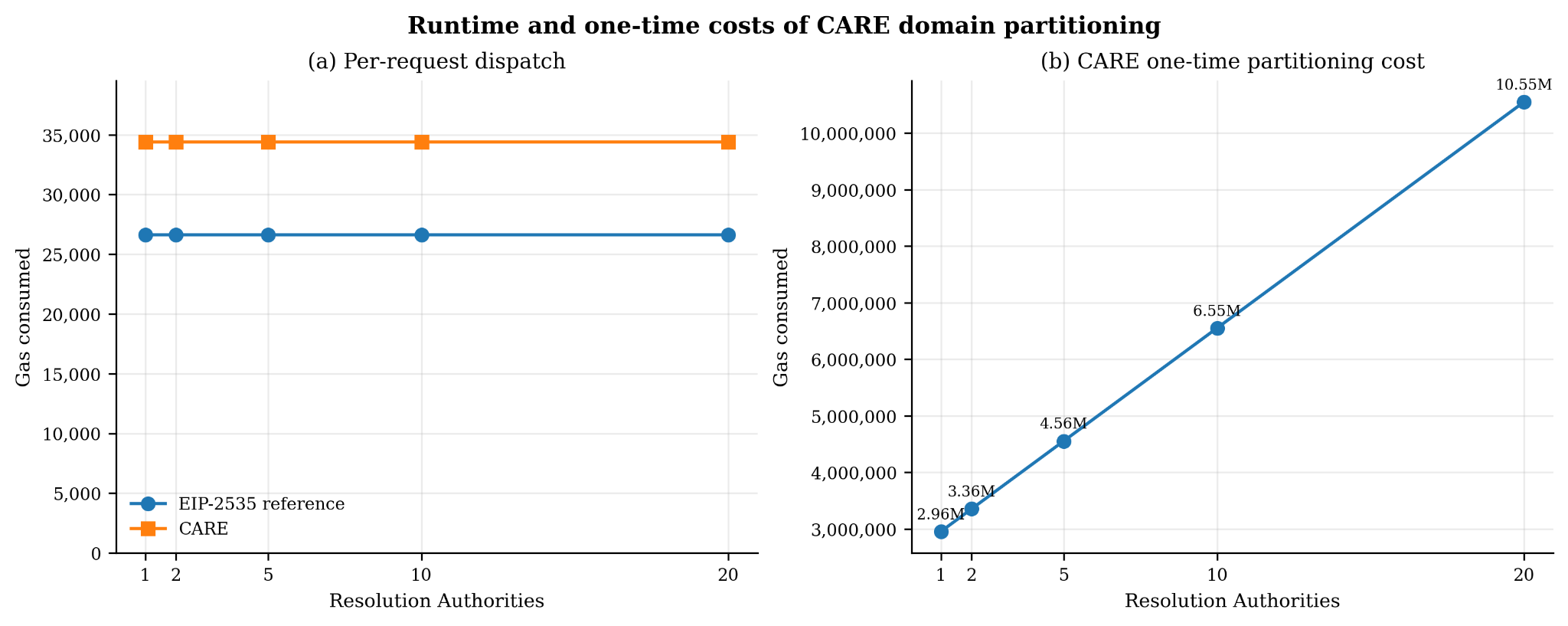}
\caption{Effect of CARE domain partitioning on per-request dispatch gas and one-time deployment and configuration cost while holding the total capability count constant at 100.}\label{fig:domain-partitioning}
\end{figure}

\subsection{RQ3 --- Localized Evolution}\label{rq3-localized-evolution}

\begin{table}[H]
\centering
\caption{Localized Capability Reconfiguration Gas Across Replacement Cardinalities}\label{tab:9}
\scriptsize
\renewcommand{\arraystretch}{1.15}
\setlength{\tabcolsep}{3pt}
\begin{tabularx}{\textwidth}{@{}>{\RaggedRight\arraybackslash}X>{\RaggedRight\arraybackslash}X>{\RaggedRight\arraybackslash}X>{\RaggedRight\arraybackslash}X>{\RaggedRight\arraybackslash}X>{\RaggedRight\arraybackslash}X@{}}
\toprule
\textbf{Replaced entries} & \textbf{EIP-2535} & \textbf{DiamondCore} & \textbf{CARE} & \textbf{CARE vs. DiamondCore} & \textbf{CARE vs. EIP-2535} \\
\midrule
1 & 134,087 & 33,261 & 33,464 & +0.61\% & $-$75.04\% \\
5 & 179,512 & 66,509 & 67,348 & +1.26\% & $-$62.48\% \\
10 & 258,112 & 108,081 & 109,715 & +1.51\% & $-$57.49\% \\
25 & 470,559 & 232,761 & 236,756 & +1.72\% & $-$49.69\% \\
\bottomrule
\end{tabularx}
\end{table}

\begin{figure}[H]
\centering
\includegraphics[width=0.96\linewidth,height=0.70\textheight,keepaspectratio]{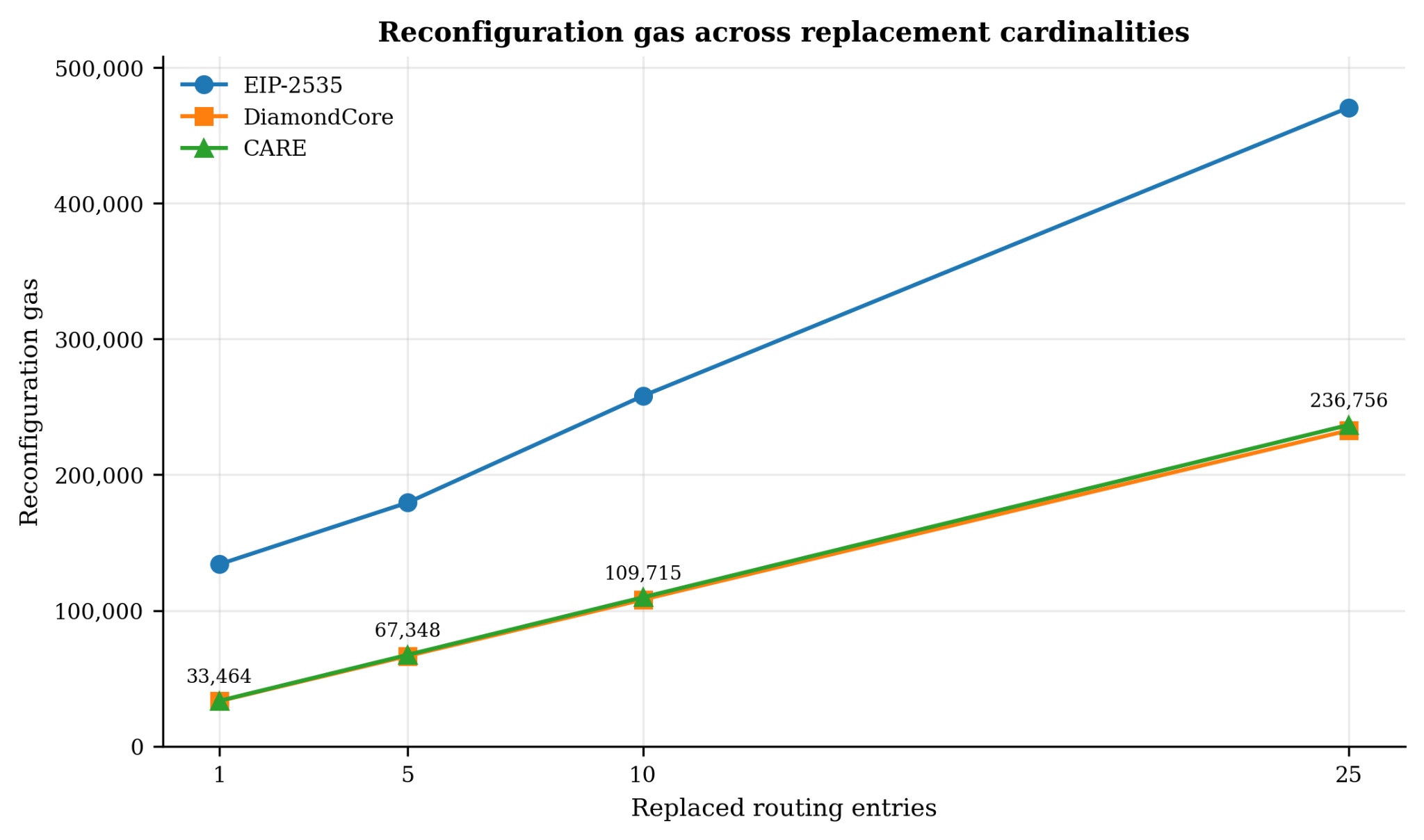}
\caption{Localized reconfiguration cost for CARE, DiamondCore, and the evaluated EIP-2535 implementation across increasing replacement cardinalities. CARE remains close to the routing-only control while the evaluated EIP-2535 path incurs additional selector and facet management cost.}\label{fig:localized-reconfiguration}
\end{figure}

CARE remains close to the routing-only control across all evaluated replacement sizes.

The difference from DiamondCore ranges from 0.61\% to 1.72\%.

The practical EIP-2535 path performs more management work and therefore exhibits a substantially different gas profile.

For one replacement:

\begin{center}
CARE = 33,464
\end{center}

versus:

\begin{center}
EIP-2535 = 134,087
\end{center}

For 25 replacements:

\begin{center}
CARE = 236,756
\end{center}

versus:

\begin{center}
EIP-2535 = 470,559
\end{center}

Opcode traces explain much of the difference. For one replacement, CARE and DiamondCore each perform one SLOAD and one SSTORE, whereas the evaluated EIP-2535 path performs 24 SLOAD and 15 SSTORE operations and includes a DELEGATECALL.

At 25 replacements, CARE and DiamondCore each perform 25 SLOAD and 25 SSTORE operations; the EIP-2535 path performs 312 SLOAD and 207 SSTORE operations.

These measurements should not be interpreted as evidence that CARE is universally cheaper than Diamond. The operations perform different amounts of bookkeeping.

The more defensible result is that CARE's localized capability update remains operationally close to a minimal routing-map update in the evaluated implementation.

This empirical result aligns with the formal property tested by LocalizedCapabilityReconfiguration: changing one domain-capability binding leaves unrelated bindings unchanged.

RQ3: localized CARE evolution remains near the minimal routing control while avoiding modification of unrelated resolution state.

\subsection{RQ4 --- Independent Architectural Concerns}\label{rq4-independent-architectural-concerns}

\begin{table}[H]
\centering
\caption{Incremental Gas Cost of CARE Architectural Concerns}\label{tab:10}
\small
\renewcommand{\arraystretch}{1.15}
\setlength{\tabcolsep}{3pt}
\begin{tabularx}{\textwidth}{@{}>{\RaggedRight\arraybackslash}X>{\RaggedRight\arraybackslash}X>{\RaggedRight\arraybackslash}X@{}}
\toprule
\textbf{Scenario} & \textbf{Gas} & \textbf{Difference from relevant base} \\
\midrule
A0 --- RK + RA + EM & 34,176 & --- \\
A1 --- A0 + SS read & 40,541 & +6,365 \\
A2 --- A0 + SA call & 40,497 & +6,321 \\
A3 --- A0 + SS + SA & 48,109 & +13,933 \\
A4 --- A3 + OH & 60,471 & +12,362 \\
A5 --- A3 + forced OH failure & 59,937 & +11,828 \\
\bottomrule
\end{tabularx}
\end{table}

SS and SA introduce similar but independently measurable overheads: 6,365 and 6,321 gas respectively.

The combined path consumes 48,109 gas, 13,933 above A0.

Adding successful observation raises the path to 60,471 gas.

In A5, the Observation Hub is forced to revert. Under the selected reference implementation policy, the observation failure is caught after successful business execution, and the transaction consumes 59,937 gas without changing the successful business result.

This behavior demonstrates only the chosen reference policy. CARE does not require observation failures to be ignored.

The result is architecturally useful because SS and SA are measured independently. Shared state ownership and shared service behavior remain separate both conceptually and operationally.

RQ4: SS, SA, and OH have distinct measurable implementation costs; CARE does not make responsibility separation operationally free.

\subsection{Empirical Summary}\label{empirical-summary}

The experiments expose three principal cost dimensions.

First, explicit domain-scoped resolution adds runtime indirection, measured at approximately 7.7k gas in the tested execution paths.

Second, finer domain decomposition increases one-time architectural setup cost even though request dispatch remained stable within the tested domain counts.

Third, localized capability replacement remains close to a minimal mapping update.

CARE therefore does not eliminate architectural cost. Instead, it redistributes cost among:

\begin{itemize}
\item
  Runtime Dispatch;
\item
  Architectural Construction;
\item
  Localized Evolution;
\end{itemize}

This cost structure is central to interpreting CARE as an architecture rather than as a gas-optimization technique.

\section{Threat Model and Security Characterization}\label{sec:security}

\subsection{Trust Assumptions}\label{trust-assumptions}

The security analysis treats the integrity of the following as security-relevant:

\begin{itemize}
\item
  RK configuration;
\item
  domain-to-RA registration;
\item
  RA capability bindings;
\item
  Shared Architectural State;
\item
  Observation Hub records;
\item
  execution through registered modules.
\end{itemize}

The configured administrator is trusted and assumed not to be compromised.

RK, RA, SS, and OH reference components are trusted.

Ordinary callers, unauthorized contracts, external dependencies, adversarial calldata, and unregistered modules are potentially untrusted.

The most consequential trust assumption concerns Execution Modules.

The Runtime Kernel invokes an EM through delegatecall. The selected module therefore executes inside RK's storage context.

Consequently:

\begin{quote}
A registered Execution Module is part of the Runtime Kernel's trusted computing base.
\end{quote}

Responsibility separation is not a bytecode sandbox.

\subsection{Configuration Integrity}\label{configuration-integrity}

The reference implementation restricts:

\begin{itemize}
\item
  domain registration;
\item
  capability binding;
\item
  Shared Architectural State mutation;
\item
  observation emitter configuration
\end{itemize}

to the configured administrative authority.

Adversarial tests attempt these operations from unauthorized accounts and require failure.

The reference implementation also validates domain--authority consistency during Resolution Authority registration.

These controls protect the evaluated configuration surface from ordinary unauthorized mutation. They do not protect against malicious or compromised administration.

\subsection{Shared-State Boundary}\label{shared-state-boundary}

The Solidity realization applies a stricter SS write policy than the general architectural model: architectural-state mutation is administrator-only.

An ordinary execution path through EM therefore cannot mutate SS through its public administrative interface.

This property applies only to CARE Shared Architectural State. It does not imply that arbitrary business state used by an application is similarly protected.

\subsection{Delegated Execution}\label{delegated-execution}

delegatecall preserves the Runtime Kernel execution context but allows the delegated bytecode to access its storage.

An adversarial Execution Module was therefore deliberately registered and executed through RK. The module directly overwrote a Runtime Kernel storage slot.

The successful corruption demonstrates the boundary intentionally:

\[
\text{Approved EM} \not\Rightarrow \text{Sandboxed EM}
\]

CARE controls which module is selected. The evaluated architecture does not establish that administrator-approved bytecode is safe.

A production system may introduce code-hash allowlists, formal audits, governance restrictions, code provenance requirements, or alternative execution boundaries. These are possible hardening mechanisms, not intrinsic CARE properties.

\subsection{Reentrancy and Nested Execution}\label{reentrancy-and-nested-execution}

CARE does not inherently prevent synchronous reentrancy.

A delegated EM can make an external call to RK and request another registered domain-capability pair. Adversarial testing demonstrates that execution initiated in one domain can synchronously trigger a second execution in another domain.

This behavior is not necessarily malicious. Cross-domain composition may be legitimate.

The important conclusion is instead:

\[
\text{Domain-Scoped Resolution} \neq \text{Runtime Isolation}
\]

Reentrancy-sensitive workflows remain responsible for appropriate implementation-level protection.

A global non-reentrancy constraint is intentionally not defined as a CARE invariant because it could prohibit valid nested capability composition.

\subsection{Observation Integrity}\label{observation-integrity}

The reference OH accepts records only from authorized emitters, and emitter configuration is administrator-controlled.

Unauthorized observation attempts and unauthorized emitter configuration are rejected by the adversarial tests.

The executeObserved reference path performs business execution before attempting the observation call and uses handled failure semantics. A normal OH revert therefore does not revert a completed business operation under this implementation policy.

This result should not be generalized to all observation failures or to all CARE implementations. Gas exhaustion, malicious dependencies, or compliance-sensitive observation may require different semantics.

\subsection{Security Interpretation}\label{security-interpretation}

The evaluated implementation therefore enforces several useful boundaries:

\begin{itemize}
\item
  unauthorized RK configuration changes are rejected;
\item
  unauthorized RA remapping is rejected;
\item
  domain--authority conformance is checked;
\item
  SS administrative mutation is protected;
\item
  observation producers are authorized.
\end{itemize}

Three important risks remain outside these controls:

\begin{enumerate}
\def\labelenumi{\arabic{enumi}.}
\item
  registered EM bytecode is trusted under delegatecall;
\item
  CARE does not inherently isolate reentrant or nested execution;
\item
  administrator compromise remains outside the evaluated protection model.
\end{enumerate}

CARE's security contribution is therefore explicit trust-boundary definition and adversarial characterization, not a claim of complete smart-contract security.

The existing artifact explicitly reaches the same conclusion: unauthorized configuration mutations are rejected, while delegated EMs remain trusted and CARE provides no generic reentrancy isolation.

\section{Discussion and Threats to Validity}\label{sec:discussion}

\subsection{Runtime Overhead and Architectural Justification}\label{runtime-overhead-and-architectural-justification}

CARE deliberately introduces an additional resolution boundary. In the evaluated implementation, each request traverses the Runtime Kernel and the domain-specific Resolution Authority before delegated business execution. This structure produces an approximately fixed 7.7k-gas premium relative to the evaluated EIP-2535 path.

Opcode traces help explain this premium. Across the evaluated workloads, CARE performs one additional SLOAD and one additional STATICCALL relative to the EIP-2535 execution path, while both paths perform one DELEGATECALL. The measured premium is therefore consistent with the additional domain-to-authority resolution step rather than with workload-specific business computation.

This overhead should be interpreted as the operational cost of representing domain-scoped resolution ownership explicitly on the EVM. CARE does not attempt to minimize transaction gas in isolation; it exchanges an additional per-request indirection for explicit resolution ownership, independently evolvable domains, and localized capability reconfiguration.

\subsection{Applicability: Runtime Cost versus Localized Evolution}\label{applicability-runtime-cost-versus-localized-evolution}

CARE pays the domain-scoped resolution premium on every execution, whereas reconfiguration cost is incurred only when capability bindings change. The lower reconfiguration cost should therefore not be interpreted as automatically amortizing the recurring runtime overhead.

CARE is most applicable to evolution-intensive modular systems in which functional domains or capabilities are expected to change independently over time. In such systems, explicit ownership of resolution and localized reconfiguration may justify the recurring dispatch cost. The case becomes stronger when independent capability evolution is a recurring architectural requirement rather than an exceptional maintenance event.

Conversely, a simple contract with limited modularity, a small and stable execution surface, or little expectation of independent evolution may not benefit from this additional indirection. Such systems may reasonably prefer direct execution, a conventional proxy, or a simpler routing mechanism.

The appropriate interpretation is therefore:

\[
\text{Additional Runtime Indirection} \longleftrightarrow \text{Localized Resolution Ownership}
\]

CARE is a design option for systems that value controlled, domain-local evolution, not a default architecture for every smart contract.

\subsection{Domain Granularity}\label{domain-granularity}

CARE does not prescribe one ideal domain size.

Coarse domains require fewer Resolution Authorities and therefore lower deployment and configuration cost.

Finer domains provide more independent resolution boundaries but require more architectural setup.

Domain selection is consequently an architectural design decision rather than a parameter that should always be maximized.

The empirical evidence only shows that domain count did not affect per-request dispatch gas from one to twenty RAs in the evaluated mapping-based implementation. It does not prove unrestricted scalability.

\subsection{CARE and EIP-2535}\label{care-and-eip-2535}

The empirical comparison with EIP-2535 requires careful interpretation.

EIP-2535 addresses selector-to-facet organization and standardized facet management. CARE asks which architectural component owns each responsibility and which domain owns capability resolution.

The two therefore operate at different abstraction levels.

A Diamond implementation could potentially adopt CARE-like responsibility boundaries. Likewise, a CARE architecture need not necessarily be implemented using the Solidity routing strategy evaluated here.

The EIP-2535 comparison is useful because it provides a mature modular execution baseline. It should not be interpreted as a claim that CARE replaces Diamond.

In particular, the lower reconfiguration gas observed for CARE reflects differences in the tested management operations. DiamondCore was included precisely to expose the cost of a simple routing update separately from full EIP-2535 bookkeeping.

\begin{table}[H]
\centering
\caption{Feature-Level Comparison of CARE and EIP-2535}\label{tab:11}
\small
\renewcommand{\arraystretch}{1.15}
\setlength{\tabcolsep}{3pt}
\begin{tabularx}{\textwidth}{@{}>{\raggedright\arraybackslash}X>{\raggedright\arraybackslash}X>{\raggedright\arraybackslash}X@{}}
\toprule
\textbf{Dimension} & \textbf{EIP-2535} & \textbf{CARE} \\
\midrule
Resolution unit & Function selector $\rightarrow$ facet & Capability + domain $\rightarrow$ EM \\
Introspection & Standardized IDiamondLoupe & Per-capability RA resolution; no standardized system-wide loupe equivalent \\
Change recording & Required DiamondCut event for function changes & CapabilityBound in the reference RA and optional architectural observation; no equivalent core-wide standardized history interface \\
Batch reconfiguration & diamondCut atomically adds/replaces/removes multiple selectors and may execute initialization & bindBatch atomically rebinds multiple capabilities within one RA; no standardized cross-domain upgrade transaction \\
Immutable functionality & Immutable functions and immutable Diamonds are supported & No core immutability/finalization mechanism is currently defined \\
Update validation & Standardized Add/Replace/Remove validity rules & Authorized domain-local rebinding with module validity checks \\
Ownership / governance & Authentication design is explicitly implementation-defined & Governance protocol is outside CARE core; evaluated implementation uses administrative authority \\
Delegated execution & Facets execute through delegatecall in Diamond storage context & EMs execute through delegatecall in RK storage context \\
Responsibility ownership & Not defined as a responsibility-oriented architecture & Explicit ownership across RK, RA, EM, SS, SA, and OH \\
\bottomrule
\end{tabularx}
\end{table}

The comparison therefore involves partially overlapping but non-identical feature sets. EIP-2535 standardizes operational mechanisms that are not currently part of the CARE core, particularly facet introspection, standardized change history, explicit Add/Replace/Remove semantics, and immutable functions. These mechanisms provide useful transparency, upgrade-safety, and tooling properties.

They should not, however, be interpreted as establishing that Diamond is inherently more secure. EIP-2535 explicitly leaves ownership and authentication design to individual implementations and notes that diamondCut can perform delegated initialization with access to Diamond storage and must therefore be carefully restricted. CARE similarly retains trusted administrative and delegated-execution boundaries. The security comparison is therefore feature- and threat-specific rather than a claim of universal superiority by either architecture.

\subsection{Construct and Internal Validity}\label{construct-and-internal-validity}

Gas accurately measures EVM execution cost under controlled conditions but is not a direct measure of maintainability, security, reliability, developer productivity, latency, or throughput.

Similarly, the five derived CARE properties describe architectural structure. They do not empirically prove higher-level software quality.

The experiments reduce internal confounding by reusing the same underlying workload across execution paths where possible. The domain-growth experiment keeps total capability count fixed. The concern-ablation experiment enables responsibilities incrementally, and DiamondCore separates mapping mutation from EIP-2535 management overhead.

Nevertheless, the compared systems are not semantically identical in every management operation. Gas differences should therefore be understood as measurements of the evaluated mechanisms, not perfect instruction-for-instruction comparisons.

\subsection{External Validity}\label{external-validity}

The Solidity implementation is a controlled research artifact rather than a production migration. The W4 marketplace workload introduces a representative ERC-20/ERC-721 application flow, but it is intentionally kept identical across execution paths so that architectural overhead remains measurable without application-specific confounding.

The workload ladder spans minimal execution, storage mutation, external access, state-changing interactions, and a marketplace-style ERC-20/ERC-721 workflow. This increases coverage but cannot represent every DeFi, DAO, gaming, identity, payment, or cross-chain system.

No production-scale migration study is included.

This choice improves isolation of the architectural variables studied here but limits claims about industrial adoption, migration effort, and long-term maintenance.

A production or mature open-source case study is therefore an important next empirical step.

The evaluation is also EVM-specific. CARE's conceptual model is not defined in Solidity terms, but no experimental evidence is currently provided for Solana, Move-based systems, TON, Hyperledger Fabric, or other execution environments.

\subsection{Formal-Validation Validity}\label{formal-validation-validity}

TLC exhaustively explores the reachable states of the selected bounded model, not every possible CARE deployment.

The absence of an invariant violation means no counterexample was found inside that explored state space.

The behavioral specification abstracts from detailed Solidity semantics including arbitrary bytecode behavior, storage layout, gas exhaustion, and complete reentrancy behavior.

The machine-checked results therefore complement rather than replace implementation-level testing.

\subsection{Security Validity}\label{security-validity}

The adversarial evaluation is not a comprehensive smart-contract audit.

It targets architectural configuration mutation, state access, observation authenticity, delegated storage behavior, and nested execution.

Economic attacks, arbitrary dependency behavior, denial-of-service strategies, malicious administration, and the full range of Solidity vulnerabilities remain outside the evaluated coverage.

This distinction is particularly important because CARE deliberately leaves registered EM bytecode within the TCB.

\subsection{Upgrade and State-Evolution Validity}\label{upgrade-and-state-evolution-validity}

Localized capability replacement is not equivalent to a complete upgrade lifecycle.

CARE currently does not define:

\begin{itemize}
\item
  version approval;
\item
  timelocks;
\item
  staged activation;
\item
  rollback;
\item
  recovery from failed upgrades;
\item
  compatibility negotiation;
\item
  schema migration;
\item
  historical-state transformation.
\end{itemize}

Independent SS ownership provides an architectural boundary within which state evolution can be managed. It does not itself solve schema compatibility.

The empirical evolution results should therefore be read narrowly: they measure localized resolution-map change, not all consequences of upgrading production smart-contract systems.

\subsection{Overall Architectural Interpretation}\label{overall-architectural-interpretation}

Taken together, the formal and empirical evidence supports a specific conclusion.

CARE introduces:

\begin{itemize}
\item
  an additional runtime resolution step;
\item
  explicit architectural responsibility boundaries;
\item
  independent domain-resolution ownership;
\item
  independent state and service ownership;
\item
  explicit observation ownership;
\item
  localized capability replacement.
\end{itemize}

The architecture does not minimize all costs. It makes several costs and ownership decisions separately identifiable.

That distinction is central.

A routing architecture may answer where execution goes. CARE attempts to additionally make explicit who owns the decision, which scope the decision belongs to, and which surrounding architectural responsibilities remain independently evolvable.

\section{Conclusion}\label{sec:conclusion}

This paper presented CARE, a responsibility-oriented architecture for modular and upgradeable smart-contract systems.

CARE starts from the observation that modular execution does not by itself establish modular architectural responsibility. A system may distribute executable code across replaceable components while leaving ownership of execution coordination, target resolution, shared state, reusable services, and observation implicit.

CARE separates these responsibilities across six architectural roles: Runtime Kernel, Resolution Authority, Execution Module, Shared Architectural State, Shared Architectural Service, and Observation Hub.

Its central mechanism is domain-scoped capability resolution:

\[
(\mathrm{Capability},\mathrm{Domain})\rightarrow \mathrm{ExecutionModule}
\]

with execution coordination owned independently by the Runtime Kernel.

A formal architectural model defines entities, responsibility ownership, legal invocation relations, state-access rules, service use, and domain-scoped resolution. Five structural properties follow from the model: unique responsibility ownership, domain-scoped centralized resolution, architectural-state ownership independence, shared-service ownership independence, and structural observation separation.

The behavioral model was additionally encoded in TLA+ and evaluated with TLC. Seven architectural invariants were checked over a bounded state space, with intermediate TLC results used to refine the architectural model before the final validation run. The final exploration generated 1,805,161 states, including 360,980 distinct reachable states, and completed without invariant violation in the evaluated model.

The Solidity evaluation demonstrated that this architectural structure has a measurable operational cost. CARE introduced an approximately fixed 7.7k-gas runtime routing premium relative to the evaluated EIP-2535 path. Increasing domain count did not increase measured request dispatch cost within the tested bounds, but finer domain decomposition substantially increased one-time architectural setup cost.

Localized evolution exhibited the opposite trade-off. Capability replacement remained within approximately 0.61--1.72\% of the routing-only control across the tested replacement sizes and required less gas than the evaluated EIP-2535 reconfiguration path. The result should not be interpreted as universal superiority over Diamond; it shows that CARE's domain-local update remains close to a minimal mapping mutation in the evaluated realization.

These results also delimit CARE's intended applicability. The architecture is most relevant to modular systems in which capabilities or functional domains are expected to evolve independently and where explicit resolution ownership is an architectural requirement. For simple or rarely evolving contracts, the recurring resolution premium may not be justified relative to a direct or simpler routing architecture.

Concern ablation further showed that Shared Architectural State, Shared Architectural Services, and Observation Hub integration each have separately measurable costs. Responsibility separation therefore has an operational price rather than being an abstract property with no runtime consequence.

Security evaluation exposed equally important boundaries. Unauthorized architectural mutation is rejected by the tested configuration controls, but registered Execution Modules remain trusted under delegatecall. Domain-scoped resolution does not provide runtime isolation, and CARE does not inherently prevent reentrancy. Administrator compromise also remains outside the evaluated protection model.

The contribution of CARE is therefore neither universally lower gas nor a universal security improvement. It is an architectural model in which responsibility ownership, resolution scope, evolution boundaries, and their operational costs can be reasoned about separately.

Future work should investigate complete upgrade governance, capability and authority versioning, rollback, state and schema migration, stronger execution isolation, and realizations outside the EVM. A complementary empirical direction is a migration study using an existing production-scale or mature open-source smart-contract system to measure applicability, decomposition effort, and long-term evolution.

CARE should consequently be understood as a foundation for responsibility-oriented smart-contract architecture rather than as a final replacement for every modularity or upgradeability mechanism.

\bibliographystyle{IEEEtran}
\bibliography{care-references}
\end{document}